\documentclass[journal,10pt]{IEEEtran}

\ifCLASSINFOpdf
\else
   \usepackage[dvips]{graphicx}
\fi
\usepackage{url}

\usepackage{amsmath,graphicx,multirow,array,cite,epstopdf,amsmath,flushend}
\begin{document}

\title{Real-time Monaural Speech Enhancement With Short-time Discrete Cosine Transform}

\author{Qinglong Li, Fei Gao, Haixin Guan, and Kaichi Ma
\thanks{The manuscript was submitted on February 8, 2021. This work was supported by the AI Labs of Unisound AI Technology Co. Ltd.}
\thanks{The authors are all with AI-Labs of Unisound AI Technology Co. Ltd. (e-mail: \{liqinglong, gaofei, guanhaixin, makaichi \}@unisound.com).}
}

%\markboth{Journal of \LaTeX\ Class Files, Vol. 14, No. 8, February 2021}
\maketitle

\begin{abstract}
Speech enhancement algorithms based on deep learning have been improved in terms of speech intelligibility and perceptual quality greatly. Many methods focus on enhancing the amplitude spectrum while reconstructing speech using the mixture phase. Since the clean phase is very important and difficult to predict, the performance of these methods will be limited. Some researchers attempted to estimate the phase spectrum directly or indirectly, but the effect is not ideal. Recently, some studies proposed the complex-valued model and achieved state-of-the-art performance, such as deep complex convolution recurrent network (DCCRN). However, the computation of the model is huge. To reduce the complexity and further improve the performance, we propose a novel method using discrete cosine transform as the input in this paper, called deep cosine transform convolutional recurrent network (DCTCRN). Experimental results show that DCTCRN achieves state-of-the-art performance both on objective and subjective metrics.  Compared with noisy mixtures, the mean opinion score (MOS) increased by 0.46 (2.86 to 3.32) absolute processed by the proposed model with only 2.86M parameters.
\end{abstract}


\IEEEpeerreviewmaketitle

\section{Introduction}
\IEEEPARstart{S}{peech} enhancement (SE) is one important topic in the signal processing area. Its goal is to extract target speech from the background noise environments to improve the intelligibility and perceptual quality. However, it is not easy to restore clean target speech perfectly especially with a single microphone. Researchers have proposed many algorithms to deal with this problem in past years. Traditional methods can deal with stationary noise effectively but powerless against non-stationary noise, such as the Wiener Filter method. Deep learning (DL) based methods make up for the above shortcomings, they regard SE as a supervised learning problem in the time-frequency (T-F) domain or in the time domain.

Typical TF domain speech enhancement methods only enhance the magnitude spectrum and use the noisy phase to reconstruct the clean target speech \cite{xu2014an}. This may because there is no clear structure in the phase spectrogram, which makes it difficult to estimate the clean phase from the noisy phase. These methods can be divided into two categories: mapping and mask-based methods. The mapping methods use a deep neural network (DNN) to learn the mapping function between the noisy magnitude spectrum and its corresponding target speech magnitude spectrum. The mask-based methods estimate a mask that classifies every portion of the signal either as speech or noise, and then by weighting, or filtering the noisy speech with this mask, the enhanced clean speech signal can be generated. Common mask functions include ideal binary mask (IBM) \cite{hu2001speech}, ideal ratio mask (IRM) \cite{srinivasan2006binary}, and spectral magnitude mask (SMM) \cite{wang2014on}, which show better performance than direct spectral mapping.

Recently, some research has shown the importance of phase for spectrograms to be resynthesized back into time-domain waveforms \cite{paliwal2011the}. Subsequently, some methods with phase information estimation were proposed. Phase-sensitive mask (PSM) \cite{erdogan2015phase-sensitive} was the first one that utilizes phase information showing the feasibility of phase estimation. Later,  the complex ratio mask (CRM) \cite{williamson2016complex} and complex spectral mapping (CSM) \cite{tan2019complex} were proposed, which can reconstruct clean speech perfectly in theory. Although the input and the mask are complex-valued, they all use a real-valued network to learn. In 2017, Chiheb \textit{et al} proposed a deep complex network (DCN) \cite{Chiheb2017complex}, which has much better performance than real-valued networks. After that, deep complex u-net (DCUNET) \cite{choi2018phase} has combined the advantages of DCN and UNET \cite{ronneberger2015u} to deal with the complex-valued spectrogram. Recently, built upon the models DCN and CRN, a deep complex convolution recurrent network (DCCRN) \cite{hu2020dccrn} was proposed, which achieved state-of-the-art performance in the TF domain. DCCRN won the championship on the real-time track of the Interspeech 2020 deep noise suppression (DNS) challenge.

Another popular method is to form an end-to-end algorithm in the time domain. In 2019, Luo \textit{et al} proposed Conv-TasNet, which is an encoder-decoder architecture \cite{luo2019conv}. It used an encoder layer to extract the input features, and multiple temporal convolutional networks (TCN) as the processing modules followed by a decoder layer to reconstruct the target speech. Although achieving the best performance, so many 1-dimension convolutional layers are adopted to extract context information leads to large time delay and computational complexity, which limits its practical usage in delay-sensitive applications.

In this paper, we propose a real-valued neural network to estimate the mask which including implicit phase information. It uses short-time discrete cosine transform (STDCT) as the input feature that has been proved helpful to the model \cite{ahmed1974discrete, dct-dl}. Experimental results show that the proposed DCTCRN model outperforms the DCCRN in terms of objective and subjective metrics while reducing the calculation greatly.

\begin{figure}[t]
\centering\includegraphics[width=8cm]{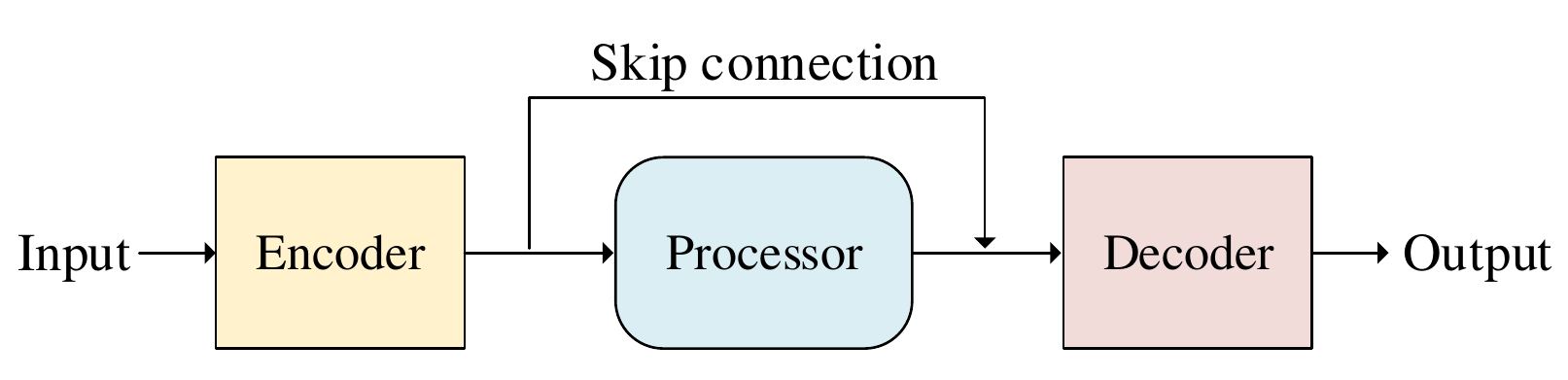}
\caption{Encoder-decoder architecture}
\label{fig:structure}
\end{figure}

\section{The DCTCRN Model}
\label{sec:dctcrn}
\subsection{Encoder-decoder architecture}

The typical encoder-decoder (ED) architecture including an encoder, a processor, and a decoder. The encoder extracts high-level features from the input to reduce the complexity, which is helpful for model training. Then, the processor uses these features to learn temporal dependencies and do the enhancement. Eventually, the decoder reconstructs the low-resolution features to the original size of the input and predicts the output targets. It is worth mentioning that, to improve flows of information and gradients throughout the network, skip connections that concatenate the output of each encoder layer to the input of each decoder layer is often utilized. The architecture is shown in Fig. \ref{fig:structure}.

CRN is one of the classical ED architecture models\cite{tan2018convolutional}. It uses multi convolutional layers as the encoder, LSTM as the processor, and transpose convolutional layers as the decoder. CRN leads to consistently better objective speech intelligibility and quality than the long short-term memory (LSTM) model. Using convolutional layers and transpose convolutional layers makes trainable parameters much fewer. This conclusion is confirmed again in the complex CRN model \cite{tan2019complex}. Combine CRN architecture and complex layer, Hu \textit{et al.} proposed DCCRN \cite{hu2020dccrn}. While achieving state-of-the-art performance in the real-time application,  the computation of complex layers is huge which leads to a very big challenge to the device.

\begin{figure*}[t]
\centering\includegraphics[width=18cm]{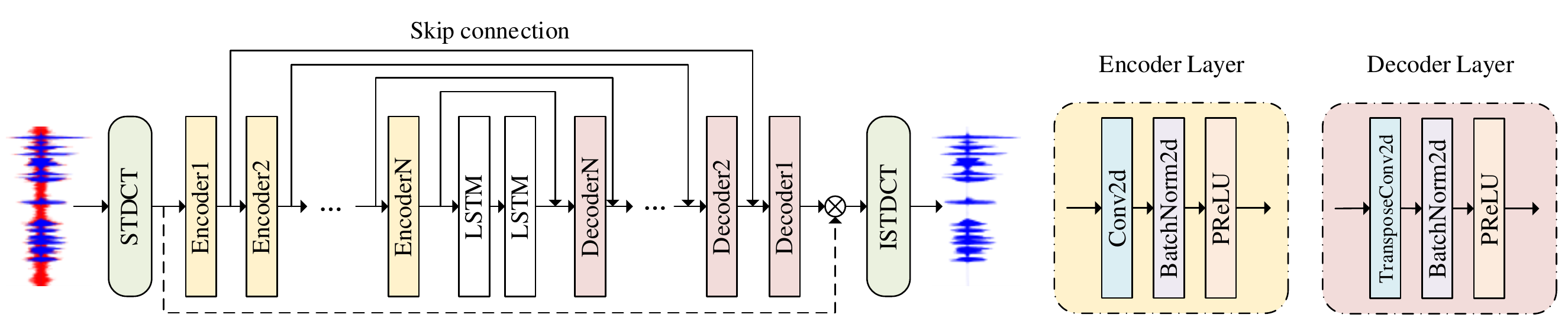}
\caption{Proposed model architecture}
\label{fig:model_structure}
\end{figure*}

In reality, the real and imaginary parts of STFT might not be necessarily as strongly correlated as we think, and the manually designed complex layer may not suitable for the real operation of the complex-value. To make up for the shortcomings of DCCRN, in this work, we propose the DCTCRN model. It replaces STFT and ISTFT with STDCT and ISTDCT, respectively, and uses a real-valued network layer to predict the mask. Unlike the explicit phase in STFT, STDCT including an implicit phase, so we can train a real-valued network without manually design complex layers. The DCTCRN model architecture is shown in Fig. \ref{fig:model_structure}.

\subsection{Input feature}

There are two main categories of input features for the SE model: the time-domain feature, the transform-domain feature, or its variants. The time-domain feature is a low-dimensional spatial feature with all information of the speech. Generally, the information of speech is very complex which is difficult to distinguish every portion if in low-dimensional space. Therefore, it is helpful to improve the enhancement performance when increasing the complexity of the network. However, the calculation amount is also increased fast at the same time, such as Conv-TasNet.

The commonly used transform-domain input feature is the discrete Fourier transform (DFT). It transforms the time-domain signal into the frequency domain which is more discriminative for every portion. And there is no information lost during the transformation. However, DFT data is complex-valued, the computation is much complexity when operating using real and imaginary parts at the same time. To reduce the computation, some researchers use the variants of STFT as input, such as amplitude/energy spectrum, Log-Mel spectrum \cite{delfarah2017features}, etc. They all enhance amplitude spectrum without estimating clean phase, this would limit the enhancement performance.

DCT is a real-valued transformation without information lost and contains implicit phase. This avoids the problem of manually design a complex network to estimate the explicit phase. Meanwhile, DCT as the input feature of the real-valued network can greatly reduce the amount of calculation. The DCT is defined as:

\begin{equation} \label{dct}
  F(\mu)=c(\mu)\sqrt{\frac 2 N} \sum_{n=0}^{N-1}f(n)\cos[\frac {\pi\mu(2n+1)} {2N}], \mu=0,1,...,N-1
\end{equation}
and inverse DCT is
\begin{equation} \label{idct}
  f(n)=\sqrt{\frac 2 N} \sum_{n=0}^{N-1}c(\mu)F(\mu)\cos[\frac {\pi\mu(2n+1)} {2N}], n=0,1,...,N-1
\end{equation}
where,
\begin{equation*}
  c(\mu)=\left\{
  \begin{aligned}
  \sqrt {\frac{1}{2}},\quad & \mu = 0 \\
  1                ,\quad & \mu = 1,2,...,N-1
  \end{aligned}
  \right.
\end{equation*}
where $f(n)$ is the time-domain vector, $N$ is the length of the vector. DCT and IDCT all operate in the real domain.

\subsection{Training target}

During the training stage, DCTCRN uses the signal approximation method to estimate the ideal cosine mask (ICM). ICM can be defined as:
\begin{equation}\label{icm}
ICM_{t,f}=\frac{S_{t,d}}{Y_{t,d}}
\end{equation}
where $S_{t,d}$ and $Y_{t,d}$ denote the STDCT spectrogram of clean and noisy speech at the $t$-th time frame and the $f$-th subband respectively. $S_{t,d}$ and $Y_{t,d} \in R$ , meaning that $ICM_{t,f}$ is in the range $(-\infty,\infty)$.

To observe the impact of mask range on model performance, we use Parametric Rectified Linear Unit (PReLU, the range is -inf to inf), Sigmoid (the range is 0 to 1), and Tanh (the range is -1 to 1) as the activation functions of the final layer respectively. We named them DCTCRN-P, DCTCRN-S, and DCTCRN-T. Specifically, the models can be represented by the following equations.

\begin{itemize}
\item DCTCRN-P:
\begin{equation}\label{icmp}
\widehat{S} = PReLU(G(Y)) * Y
\end{equation}
\item DCTCRN-S:
\begin{equation}\label{icms}
\widehat{S} = Sigmoid(G(Y)) * Y
\end{equation}
\item DCTCRN-T:
\begin{equation}\label{icmt}
\widehat{S} = Tanh(G(Y)) * Y
\end{equation}
\end{itemize}
where $G(\cdot)$ represents the DCTCRN model function, $Y$ is the STDCT data of mixture, and $\widehat{S}$ denotes the estimated STDCT of clean speech.

Because the range of PReLU is $(-\infty,+\infty)$, we do some post-processing on the $\widehat{S}$ to ensure the amplitude of reconstructed speech is not clipped. The post-processing is formed as follows:
\begin{equation}
\widehat{S_{t,f}}=\left\{
\begin{aligned}
& \widehat{S_{t,f}},& if & \quad |\widehat{S_{t,f}}| \leq |Y_{t,f}| \\
& sign(\widehat{S_{t,f}})*|Y_{t,f}|,& if & \quad |\widehat{S_{t,f}}| > |Y_{t,f}|
\end{aligned}
\right.
\end{equation}
where $sign(\cdot)$ denotes take the symbol of elements, $|\cdot|$ means take the absolute value. We can get the estimated time-domain speech by using ISTDCT and overlap and add operations.

\subsection{Loss function}

It does not match the characteristics of human hearing when using mean squared error (MSE) as the loss function \cite{shivakumar2016perception, liu2017perceptually, martin2018deep, kolbaek2018monaural}. Recently, scale-invariant signal noise ratio (SI-SNR) has been commonly used as an evaluation metric \cite{le2019sdr}, which is used as the loss function to optimize the model in this work. SI-SNR is defined as:
\begin{equation}\label{sisnr}
\left\{
\begin{aligned}
& s_{target}=\frac{<\widehat{s},s>\cdot s}{\|s\|_2^2} \\
& e_{noise}=\widehat{s} - s \\
& SI-SNR = 10*log_{10}(\frac{\|s_{target}\|_2^2}{\|e_{noise}\|_2^2})
\end{aligned}
\right.
\end{equation}
where $s$ and $\widehat{s}$ are the clean and estimated time-domain speech data, respectively. $<\cdot, \cdot>$ denotes the dot product between two vectors and $||\cdot||_2$ is Euclidean norm (L2 norm).

\section{Experiments}
\label{sec:experiments}
\subsection{datasets}

We use the ICASSP2021 DNS Challenge dataset to the model training \cite{reddy2020icassp}, clean speech in training set is total 760.53 hours: read speech (562.72 hours), singing voice (8.80 hours), emotion data(3.6hours), Chinese mandarin data (185.41 hours). The details about the clean and noisy dataset are described in \cite{reddy2020interspeechresults}. To make full use of these data,  we generate the noisy clips with dynamic mixing during model training. In detail, during each training epoch, we convolve speech and noise with different room impulse response (RIR) randomly-selected from the DNS RIR-dataset and then simulate noisy audios dynamically by mixing reverb speech and noise at specific SNR. The SNR is randomly selected from -10dB to 20dB. The development test set in DNS is used to select the best-performed model.

During the test stage, we use the image method \cite{allen1979image} to generate 10000 simulated RIRs as the test RIR set. The room size is set to 5m×4m×3.5m with T60 range is 0.1:0.1:0.5. The locations of the microphone and speaker are randomly in the room with the height range is 1m to 1.5m. We limit the distance of the mic and speaker to 0.2m to 3m. The TIMIT corpus \cite{garofolo1993darpa} is selected as the test clean speech, NOISEX-92 \cite{varga1993assessment}, and the real-life record noise dataset as the test noise. There are nine common noises in the real-life record noise set: cafeteria, crossroad, background music, songs, public places, car inside, office, white, and train inside. We generate two test sets: reverberant and non-reverberant test set. For the reverberant test set, we first convolve every speech utterance with an RIR randomly selected from the test RIR set. And then mix the non- and reverberant speech with one noise respectively at every SNR (-6dB, -3dB, 0dB, 3dB, 6dB).

\subsection{model setup and baseline}

In this work, all the waveforms are resampled at 16kHz. We use the Hanning window of size 512 with the hop time of 8ms. The optimizer is Adam gradient \cite{kingma2014adam}. The initial learning rate is set to 0.001 for the first 100 epochs, and it will decay 0.5 when the validation loss goes up. The number of epochs is set to 300, and we use early stopping to select the best models.

The proposed DCTCRN model is compared with DCCRN-E, which is the champion model of the Interspeech2020 DNS Challenge in the real-time track and ranked second in the non-real-time track \cite{reddy2020interspeechresults}. The setups of DCCRN-E is the same as \cite{hu2020dccrn}. One frame after STFT is a 512-dimensional conjugate symmetric complex vector, DCCRN-E uses the last 256 points as the input after removing the direct current component. The input feature of DCTCRN is a 512-dimensional non-symmetrical real-valued vector. The number of encoder layer channels of DCTCRN is \{8, 16, 32, 64, 128, 128, 256\} and \{128, 128, 64, 32, 16, 8, 1\} for decoder layer channels. The kernel size and stride are set to (5,2) and (2,1) for all encoder and decoder layers, respectively. The number of two-layer LSTM nodes is set to 256.

To meet the delay requirements of the ICASSP2021 DNS Challenge, we implement a causal system. Firstly, we pad one zero-frame in front of each convolutional encoder layer at the time dimension and remove the last time frame at each transpose convolutional decoder. We use a frame length of 32ms with a stride of 8ms resulting in an algorithmic 40ms delay to satisfy the latency requirements, the limitation of the DNS Challenge is within 40ms.

\subsection{experimental results}

In this study, we use the perceptual evaluation of speech quality (PESQ)\footnote{https://github.com/vBaiCai/python-pesq}, the short-time objective intelligibility (STOI)\footnote{https://github.com/mpariente/pystoi}, and the SNR as the objective  metrics. Table \ref{tab:pesq}, Table \ref{tab:stoi}, Table \ref{tab:snr} show the objective results on the test set without reverberation, and Table \ref{tab:pesqr}, Table \ref{tab:stoir}, Table \ref{tab:snrr} are the results under reverberant conditions, respectively. In each case, the best result is highlighted by a boldface number.

\begin{table}[t]  %%%%%%%%%%%%%%%% PESQ
\centering
\small
\caption{PESQ on the non-reverberation test sets}
\label{tab:pesq}
\setlength{\tabcolsep}{1.7mm}
\begin{tabular}{c<{\centering}|c<{\centering}c<{\centering}c<{\centering}c<{\centering}c<{\centering}c<{\centering}c<{\centering}c<{\centering}c<{\centering}c<{\centering}|c<{\centering}c<{\centering}c<{\centering}c<{\centering}c<{\centering}c<{\centering}c<{\centering}c<{\centering}c<{\centering}c<{\centering}}
\hline
\hline
\textbf{test SNR} & \textbf{-6dB }& \textbf{-3dB} & \textbf{0dB} & \textbf{3dB} & \textbf{6dB} & \textbf{Avg.} \\
\hline
\hline
\textbf{noisy}     & 1.51 & 1.70 & 1.89 & 2.09 & 2.30 & 1.90 \\
\textbf{DCCRN-E}   & 2.25 & 2.54 & 2.82 & 3.06 & 3.27 & 2.79 \\
\textbf{DCTCRN-P}  & \textbf{2.31} & \textbf{2.59} & \textbf{2.85} & \textbf{3.08} & 3.28 & \textbf{2.82} \\
\textbf{DCTCRN-S}  & 2.25 & 2.53 & 2.78 & 3.03 & 3.23 & 2.77 \\
\textbf{DCTCRN-T}  & 2.30 & \textbf{2.59} & \textbf{2.85} & \textbf{3.08} & \textbf{3.29} & \textbf{2.82} \\
\hline
\hline
\end{tabular}
\end{table}

\begin{table}[t]  %%%%%%%%%%%%%%%% stoi
\centering
\small
\caption{STOI (in \%) on the non-reverberation test sets}
\label{tab:stoi}
\setlength{\tabcolsep}{1.7mm}
\begin{tabular}{c<{\centering}|c<{\centering}c<{\centering}c<{\centering}c<{\centering}c<{\centering}c<{\centering}c<{\centering}c<{\centering}c<{\centering}c<{\centering}|c<{\centering}c<{\centering}c<{\centering}c<{\centering}c<{\centering}c<{\centering}c<{\centering}c<{\centering}c<{\centering}c<{\centering}}
\hline
\hline
\textbf{test SNR} & \textbf{-6dB} & \textbf{-3dB} & \textbf{0dB} & \textbf{3dB} & \textbf{6dB} & \textbf{Avg.} \\
\hline
\hline
\textbf{noisy}     & 63.39 & 69.48 & 75.43 & 81.17 & 86.16 & 75.12 \\
\textbf{DCCRN-E}   & 76.91 & 83.50 & 88.65 & 92.13 & 94.63 & 87.16 \\
\textbf{DCTCRN-P}  & \textbf{77.88} & 84.04 & \textbf{88.94} & \textbf{92.20} & 94.66 & \textbf{87.55} \\
\textbf{DCTCRN-S}  & 77.28 & 83.57 & 88.48 & 91.97 & 94.48 & 87.16 \\
\textbf{DCTCRN-T}  & 77.84 & \textbf{84.08} & 88.87 & 92.18 & \textbf{94.67} & 87.53 \\
\hline
\hline
\end{tabular}
\end{table}

\begin{table}[!t]  %%%%%%%%%%%%%%%% snr
\centering
\small
\caption{SNR on the non-reverberation test sets}
\label{tab:snr}
\setlength{\tabcolsep}{1.7mm}
\begin{tabular}{c<{\centering}|c<{\centering}c<{\centering}c<{\centering}c<{\centering}c<{\centering}c<{\centering}c<{\centering}c<{\centering}c<{\centering}c<{\centering}|c<{\centering}c<{\centering}c<{\centering}c<{\centering}c<{\centering}c<{\centering}c<{\centering}c<{\centering}c<{\centering}c<{\centering}}
\hline
\hline
\textbf{test SNR} & \textbf{-6dB} & \textbf{-3dB} & \textbf{0dB} & \textbf{3dB} & \textbf{6dB} & \textbf{Avg.} \\
\hline
\hline
\textbf{noisy}    & -5.92 & -2.92 & 0.08 & 3.08 & 6.08 & 0.08 \\
\textbf{DCCRN-E}  & 4.45 & 7.30 & 9.92 & 12.08 & 14.29 & 9.61 \\
\textbf{DCTCRN-P} & \textbf{5.93} & \textbf{8.51} & \textbf{10.98} & \textbf{13.09} & \textbf{15.30} & \textbf{10.76} \\
\textbf{DCTCRN-S} & 5.65 & 8.28 & 10.71 & 12.91 & 15.10 & 10.53 \\
\textbf{DCTCRN-T} & 5.77 & 8.43 & 10.90 & 13.04 & 15.26 & 10.68 \\
\hline
\hline
\end{tabular}
\end{table}

From the results on the non-reverberation sets, we can find that DCTCRN-P and DCTCRN-T outperform DCCRN-E in all metrics. DCTCRN-P achieves state-of-the-art performance and DCTCRN-T is very close to it. DCTCRN-S and DCCRN-E yield similar PESQ and STOI scores, but DCTCRN-S gains almost 1dB more than DCCRN-E of SNR metric.  Obviously, the range of PReLU is the closest to that of ICM, the network can estimate the mask more accurately. The more in line with the scope of ICM value, the better the speech enhancement performance.

On the reverberant test sets, DCTCRN-T gets the best results of all conditions. It is more difficult to learn ICM in reverberant scenes, Tanh as activation limits the mask range to (-1,1) to reduce the difficulty of learning. DCTCRN-P performance is very close to DCTCRN-T. All DCTCRN models are much better than DCCRN-E on the same training and test conditions. Not only that, the proposed model has fewer parameters and computations. We use flops-counter.pytorch\footnote{https://github.com/sovrasov/flops-counter.pytorch} to compute the floating-point operations (FLOPs) and parameters of the models. The results show that the calculation of DCCRN is almost three times that of the proposed DCTCRN (DCCRN: 120.01M FLOPs, DCTCRN: 41.20M FLOPs). Moreover, the comparison of the parameters of DCCRN and DCTCRN is 3.98M and 2.86M.

To evaluate the subjective metric, we participated in track-1 in the 2nd DNS-challenge and get the results of the final MOS \cite{reddy2020icassp}. Compared with noisy mixtures, the MOS increased by 0.46 (2.86 to 3.32) absolute processed by the proposed model. During the test stage, we also discovered that the DCTCRN-T model is more robust to speech and noise than the others. This is because the mask with a limited range is easier to learn and the range of (-1,1) is close to the characteristics of the ICM.

\begin{table}[t]  %%%%%%%%%%%%%%%% PESQ
\centering
\small
\caption{PESQ on the reverberation test sets}
\label{tab:pesqr}
\setlength{\tabcolsep}{1.7mm}
\begin{tabular}{c<{\centering}|c<{\centering}c<{\centering}c<{\centering}c<{\centering}c<{\centering}c<{\centering}c<{\centering}c<{\centering}c<{\centering}c<{\centering}|c<{\centering}c<{\centering}c<{\centering}c<{\centering}c<{\centering}c<{\centering}c<{\centering}c<{\centering}c<{\centering}c<{\centering}}
\hline
\hline
\textbf{test SNR} & \textbf{-6dB }& \textbf{-3dB} & \textbf{0dB} & \textbf{3dB} & \textbf{6dB} & \textbf{Avg.} \\
\hline
\hline
\textbf{noisy}     & 1.51 & 1.73 & 1.93 & 2.15 & 2.33 & 1.93 \\
\textbf{DCCRN-E}   & 1.95 & 2.16 & 2.36 & 2.53 & 2.64 & 2.33 \\
\textbf{DCTCRN-P}  & \textbf{2.08} & 2.31 & 2.54 & 2.77 & 2.96 & 2.53 \\
\textbf{DCTCRN-S}  & 2.04 & 2.27 & 2.50 & 2.72 & 2.91 & 2.49 \\
\textbf{DCTCRN-T}  & \textbf{2.08} & \textbf{2.33} & \textbf{2.56} & \textbf{2.80} & \textbf{2.99} & \textbf{2.55} \\
\hline
\hline
\end{tabular}
\end{table}

\begin{table}[!t]  %%%%%%%%%%%%%%%% stoi
\centering
\small
\caption{STOI (in \%) on the reverberation test sets}
\label{tab:stoir}
\setlength{\tabcolsep}{1.7mm}
\begin{tabular}{c<{\centering}|c<{\centering}c<{\centering}c<{\centering}c<{\centering}c<{\centering}c<{\centering}c<{\centering}c<{\centering}c<{\centering}c<{\centering}|c<{\centering}c<{\centering}c<{\centering}c<{\centering}c<{\centering}c<{\centering}c<{\centering}c<{\centering}c<{\centering}c<{\centering}}
\hline
\hline
\textbf{test SNR} & \textbf{-6dB} & \textbf{-3dB} & \textbf{0dB} & \textbf{3dB} & \textbf{6dB} & \textbf{Avg.} \\
\hline
\hline
\textbf{noisy}     & 51.35 & 58.65 & 66.33 & 74.11 & 80.30 & 66.15 \\
\textbf{DCCRN-E}   & 64.86 & 71.76 & 77.23 & 81.80 & 84.53 & 76.04 \\
\textbf{DCTCRN-P}  & 65.87 & 73.12 & 79.21 & 84.81 & 88.69 & 78.34 \\
\textbf{DCTCRN-S}  & 65.35 & 72.74 & 78.89 & 84.49 & 88.03 & 77.90 \\
\textbf{DCTCRN-T}  & \textbf{66.42} & \textbf{73.80} & \textbf{80.01} & \textbf{85.68} & \textbf{89.35} & \textbf{79.05} \\
\hline
\hline
\end{tabular}
\end{table}

\begin{table}[!t]  %%%%%%%%%%%%%%%% snr
\centering
\small
\caption{SNR on the reverberation test sets}
\label{tab:snrr}
\setlength{\tabcolsep}{1.7mm}
\begin{tabular}{c<{\centering}|c<{\centering}c<{\centering}c<{\centering}c<{\centering}c<{\centering}c<{\centering}c<{\centering}c<{\centering}c<{\centering}c<{\centering}|c<{\centering}c<{\centering}c<{\centering}c<{\centering}c<{\centering}c<{\centering}c<{\centering}c<{\centering}c<{\centering}c<{\centering}}
\hline
\hline
\textbf{test SNR} & \textbf{-6dB} & \textbf{-3dB} & \textbf{0dB} & \textbf{3dB} & \textbf{6dB} & \textbf{Avg.} \\
\hline
\hline
\textbf{noisy}    & -6.00 & -2.99 & 0.00 & 3.01 & 6.00 & 0.00 \\
\textbf{DCCRN-E}  & 1.27 & 3.68 & 5.49 & 6.95 & 8.05 & 5.09 \\
\textbf{DCTCRN-P} & 2.65 & 5.20 & \textbf{7.44} & 9.81 & 11.97 & 7.41 \\
\textbf{DCTCRN-S} & 2.13 & 4.86 & 7.13 & 9.49 & 11.54 & 7.03 \\
\textbf{DCTCRN-T} & \textbf{2.73} & \textbf{5.21} & \textbf{7.44} & \textbf{9.88} & \textbf{12.02} & \textbf{7.46} \\
\hline
\hline
\end{tabular}
\end{table}

\section{Conclusion}
\label{sec:conclusion}

In this work, we proposed a robust deep cosine transform convolutional recurrent network for real-time speech enhancement. Experimental results show that the proposed model has fewer parameters, it greatly reduces the calculations while improving performance comparing with DCCRN. In the future, we will try to deploy DCTCRN in low computational devices.

\vfill\pagebreak
\bibliographystyle{IEEEtran}

\bibliography{reference}

\begin{thebibliography}{10}
\providecommand{\url}[1]{#1}
\csname url@samestyle\endcsname
\providecommand{\newblock}{\relax}
\providecommand{\bibinfo}[2]{#2}
\providecommand{\BIBentrySTDinterwordspacing}{\spaceskip=0pt\relax}
\providecommand{\BIBentryALTinterwordstretchfactor}{4}
\providecommand{\BIBentryALTinterwordspacing}{\spaceskip=\fontdimen2\font plus
\BIBentryALTinterwordstretchfactor\fontdimen3\font minus
  \fontdimen4\font\relax}
\providecommand{\BIBforeignlanguage}[2]{{%
\expandafter\ifx\csname l@#1\endcsname\relax
\typeout{** WARNING: IEEEtran.bst: No hyphenation pattern has been}%
\typeout{** loaded for the language `#1'. Using the pattern for}%
\typeout{** the default language instead.}%
\else
\language=\csname l@#1\endcsname
\fi
#2}}
\providecommand{\BIBdecl}{\relax}
\BIBdecl

\bibitem{xu2014an}
Y.~Xu, J.~Du, L.-R. Dai, and C.-H. Lee, ``An experimental study on speech
  enhancement based on deep neural networks,'' \emph{IEEE Signal processing
  letters}, vol.~21, no.~1, pp. 65--68, 2013.

\bibitem{hu2001speech}
G.~Hu and D.~Wang, ``Speech segregation based on pitch tracking and amplitude
  modulation,'' in \emph{Proceedings of the 2001 IEEE Workshop on the
  Applications of Signal Processing to Audio and Acoustics (Cat. No.
  01TH8575)}.\hskip 1em plus 0.5em minus 0.4em\relax IEEE, 2001, pp. 79--82.

\bibitem{srinivasan2006binary}
S.~Srinivasan, N.~Roman, and D.~Wang, ``Binary and ratio time-frequency masks
  for robust speech recognition,'' \emph{Speech Communication}, vol.~48,
  no.~11, pp. 1486--1501, 2006.

\bibitem{wang2014on}
Y.~Wang, A.~Narayanan, and D.~Wang, ``On training targets for supervised speech
  separation,'' \emph{IEEE/ACM transactions on audio, speech, and language
  processing}, vol.~22, no.~12, pp. 1849--1858, 2014.

\bibitem{paliwal2011the}
K.~Paliwal, K.~W{\'o}jcicki, and B.~Shannon, ``The importance of phase in
  speech enhancement,'' \emph{speech communication}, vol.~53, no.~4, pp.
  465--494, 2011.

\bibitem{erdogan2015phase-sensitive}
H.~Erdogan, J.~R. Hershey, S.~Watanabe, and J.~Le~Roux, ``Phase-sensitive and
  recognition-boosted speech separation using deep recurrent neural networks,''
  in \emph{2015 IEEE International Conference on Acoustics, Speech and Signal
  Processing (ICASSP)}.\hskip 1em plus 0.5em minus 0.4em\relax IEEE, 2015, pp.
  708--712.

\bibitem{williamson2016complex}
D.~S. Williamson, Y.~Wang, and D.~Wang, ``Complex ratio masking for monaural
  speech separation,'' \emph{IEEE/ACM transactions on audio, speech, and
  language processing}, vol.~24, no.~3, pp. 483--492, 2015.

\bibitem{tan2019complex}
K.~Tan and D.~Wang, ``Complex spectral mapping with a convolutional recurrent
  network for monaural speech enhancement,'' in \emph{ICASSP 2019-2019 IEEE
  International Conference on Acoustics, Speech and Signal Processing
  (ICASSP)}.\hskip 1em plus 0.5em minus 0.4em\relax IEEE, 2019, pp. 6865--6869.

\bibitem{Chiheb2017complex}
O.~B. Y.~Zhang D. Serdyuk S. Subramanian J. F. Santos S. Mehri N. Rostamzadeh
  Y. Bengio C.~Trabelsi and C.~J. Pal, ``Deep complex networks,''
  \emph{arXiv:1705.09792}, 2017.

\bibitem{choi2018phase}
H.-S. Choi, J.-H. Kim, J.~Huh, A.~Kim, J.-W. Ha, and K.~Lee, ``Phase-aware
  speech enhancement with deep complex u-net,'' in \emph{International
  Conference on Learning Representations}, 2018.

\bibitem{ronneberger2015u}
O.~Ronneberger, P.~Fischer, and T.~Brox, ``U-net: Convolutional networks for
  biomedical image segmentation,'' in \emph{International Conference on Medical
  image computing and computer-assisted intervention}.\hskip 1em plus 0.5em
  minus 0.4em\relax Springer, 2015, pp. 234--241.

\bibitem{hu2020dccrn}
S.~B. L. M. T. X. S. M. Z. Y. H. F. J. W. B. H.~Z. Y.~X.~Hu, Y.~Liu and L.~Xie,
  ``Dccrn: Deep complex convolution recurrent network for phase-aware speech
  enhancement,'' in \emph{Interspeech}, 2020, pp. 2472--2476.

\bibitem{luo2019conv}
Y.~Luo and N.~Mesgarani, ``Conv-tasnet: Surpassing ideal time--frequency
  magnitude masking for speech separation,'' \emph{IEEE/ACM transactions on
  audio, speech, and language processing}, vol.~27, no.~8, pp. 1256--1266,
  2019.

\bibitem{ahmed1974discrete}
N.~Ahmed, T.~Natarajan, and K.~R. Rao, ``Discrete cosine transform,''
  \emph{IEEE transactions on Computers}, vol. 100, no.~1, pp. 90--93, 1974.

\bibitem{dct-dl}
C.~Geng and L.~Wang, ``End-to-end speech enhancement based on discrete cosine
  transform,'' in \emph{2020 IEEE International Conference on Artificial
  Intelligence and Computer Applications (ICAICA)}.\hskip 1em plus 0.5em minus
  0.4em\relax IEEE, 2020, pp. 379--383.

\bibitem{tan2018convolutional}
K.~Tan and D.~Wang, ``A convolutional recurrent neural network for real-time
  speech enhancement.'' in \emph{Interspeech}, 2018, pp. 3229--3233.

\bibitem{delfarah2017features}
M.~Delfarah and D.~Wang, ``Features for masking-based monaural speech
  separation in reverberant conditions,'' \emph{IEEE/ACM Transactions on Audio,
  Speech, and Language Processing}, vol.~25, no.~5, pp. 1085--1094, 2017.

\bibitem{shivakumar2016perception}
P.~G. Shivakumar and P.~G. Georgiou, ``Perception optimized deep denoising
  autoencoders for speech enhancement.'' in \emph{Interspeech}, 2016, pp.
  3743--3747.

\bibitem{liu2017perceptually}
Q.~Liu, W.~Wang, P.~J. Jackson, and Y.~Tang, ``A perceptually-weighted deep
  neural network for monaural speech enhancement in various background noise
  conditions,'' in \emph{2017 25th European Signal Processing Conference
  (EUSIPCO)}.\hskip 1em plus 0.5em minus 0.4em\relax IEEE, 2017, pp.
  1270--1274.

\bibitem{martin2018deep}
J.~M. Martin-Donas, A.~M. Gomez, J.~A. Gonzalez, and A.~M. Peinado, ``A deep
  learning loss function based on the perceptual evaluation of the speech
  quality,'' \emph{IEEE Signal processing letters}, vol.~25, no.~11, pp.
  1680--1684, 2018.

\bibitem{kolbaek2018monaural}
M.~Kolb{\ae}k, Z.-H. Tan, and J.~Jensen, ``Monaural speech enhancement using
  deep neural networks by maximizing a short-time objective intelligibility
  measure,'' in \emph{2018 IEEE International Conference on Acoustics, Speech
  and Signal Processing (ICASSP)}.\hskip 1em plus 0.5em minus 0.4em\relax IEEE,
  2018, pp. 5059--5063.

\bibitem{le2019sdr}
J.~Le~Roux, S.~Wisdom, H.~Erdogan, and J.~R. Hershey, ``Sdr--half-baked or well
  done?'' in \emph{ICASSP 2019-2019 IEEE International Conference on Acoustics,
  Speech and Signal Processing (ICASSP)}.\hskip 1em plus 0.5em minus
  0.4em\relax IEEE, 2019, pp. 626--630.

\bibitem{reddy2020icassp}
C.~K. Reddy, H.~Dubey, V.~Gopal, R.~Cutler, S.~Braun, H.~Gamper, R.~Aichner,
  and S.~Srinivasan, ``Icassp 2021 deep noise suppression challenge,''
  \emph{arXiv preprint arXiv:2009.06122}, 2020.

\bibitem{reddy2020interspeechresults}
C.~K. Reddy, E.~Beyrami, H.~Dubey, V.~Gopal, R.~Cheng, R.~Cutler,
  S.~Matusevych, R.~Aichner, A.~Aazami, S.~Braun \emph{et~al.}, ``The
  interspeech 2020 deep noise suppression challenge: Datasets, subjective
  speech quality and testing framework,'' \emph{arXiv preprint
  arXiv:2001.08662}, 2020.

\bibitem{allen1979image}
J.~B. Allen and D.~A. Berkley, ``Image method for efficiently simulating
  small-room acoustics,'' \emph{The Journal of the Acoustical Society of
  America}, vol.~65, no.~4, pp. 943--950, 1979.

\bibitem{garofolo1993darpa}
J.~S. Garofolo, L.~F. Lamel, W.~M. Fisher, J.~G. Fiscus, and D.~S. Pallett,
  ``Darpa timit acoustic-phonetic continous speech corpus cd-rom. nist speech
  disc 1-1.1,'' \emph{NASA STI/Recon technical report n}, vol.~93, p. 27403,
  1993.

\bibitem{varga1993assessment}
A.~Varga and H.~J. Steeneken, ``Assessment for automatic speech recognition:
  Ii. noisex-92: A database and an experiment to study the effect of additive
  noise on speech recognition systems,'' \emph{Speech communication}, vol.~12,
  no.~3, pp. 247--251, 1993.

\bibitem{kingma2014adam}
D.~P. Kingma and J.~Ba, ``Adam: A method for stochastic optimization,''
  \emph{arXiv preprint arXiv:1412.6980}, 2014.

\end{thebibliography}

\end{document}